\documentclass[preprintnumbers, floatfix, 
preprintnumbers, letterpaper, superscriptaddress,nofootinbib]{revtex4}

\usepackage{graphicx}
\usepackage{microtype}
\usepackage{amsmath}
\usepackage{amssymb}
\usepackage{subfigure}
\usepackage{hyperref}
\usepackage{url}
\usepackage{xcolor}
\usepackage{color}
\usepackage{mathrsfs}
\usepackage{amsfonts}
\usepackage{latexsym}
\usepackage{epsfig}

\definecolor{vividviolet}{rgb}{0.62, 0.0, 1.0}
\definecolor{amaranth}{rgb}{0.9, 0.17, 0.31}
\definecolor{palatinateblue}{rgb}{0.15, 0.23, 0.89}
\definecolor{brightpink}{rgb}{1.0, 0.0, 0.5}
\definecolor{cornflowerblue}{rgb}{0.39, 0.58, 0.93}
\definecolor{deepcarminepink}{rgb}{0.94, 0.19, 0.22}
\definecolor{radicalred}{rgb}{1.0, 0.21, 0.37}

\hypersetup{ linktoc=all,
    colorlinks, linkcolor={palatinateblue},
    citecolor={brightpink}, urlcolor={amaranth}
}

\graphicspath{{Images/}}

\renewcommand{\d}[1]{\ensuremath{\operatorname{d}\!{#1}}}

\DeclareSymbolFont{extraup}{U}{zavm}{m}{n}
\DeclareMathSymbol{\varheart}{\mathalpha}{extraup}{86}
\DeclareMathSymbol{\vardiamond}{\mathalpha}{extraup}{87}
\makeatletter
\renewcommand*{\@fnsymbol}[1]{\ensuremath{\ifcase#1\or \clubsuit \or \vardiamond \or \varheart\or
    \spadesuit\or \mathparagraph\or \|\or **\or \dagger\dagger
    \or \ddagger\ddagger \else\@ctrerr\fi}}
\makeatother

\begin{document}

\title{Stephani Cosmology: Entropically Viable But Observationally Challenged}

\author{Yen Chin \surname{Ong}}
\email{ycong@yzu.edu.cn}
\affiliation{Center for Gravitation and Cosmology, College of Physical Science and Technology, Yangzhou University, Yangzhou 225009, China}
\affiliation{School of Physics and Astronomy,
Shanghai Jiao Tong University, Shanghai 200240, China}

\author{S. Sedigheh \surname{Hashemi}}
\email{se\_hashemi@sbu.ac.ir}
\affiliation{Department of Physics, Shahid Beheshti University, G.C., Evin, Tehran 19839, Iran}
\affiliation{School of Physics and Astronomy,
Shanghai Jiao Tong University, Shanghai 200240, China}

\author{Rui \surname{An}}
\email{an\_rui@sjtu.edu.cn}
\affiliation{School of Physics and Astronomy,
Shanghai Jiao Tong University, Shanghai 200240, China}
\affiliation{Collaborative Innovation Center of IFSA (CICIFSA),
Shanghai Jiao Tong University, Shanghai 200240, China}

\author{Bin \surname{Wang}}
\email{wang\_b@sjtu.edu.cn}
\affiliation{Center for Gravitation and Cosmology, College of Physical Science and Technology, Yangzhou University, Yangzhou 225009, China}
\affiliation{School of Physics and Astronomy,
Shanghai Jiao Tong University, Shanghai 200240, China}

\begin{abstract}
Inhomogeneous cosmological models such as the Stephani universes could, in principle, provide an explanation for the observed accelerated expansion of the Universe. 
Working with a concrete, popular model of the Stephani cosmology -- the Stephani-D\c{a}browski model, we found that it is entropically viable. We also comment on the energy conditions and the two-sheeted geometry of the spacetime. However, similar to the LTB models, despite satisfying the holographic principle, Stephani cosmology has difficulty satisfying all the constraints from observations.
\end{abstract}

\maketitle

\section{Introduction: Inhomogeneous Cosmologies and Their Constraints}\label{I}

One of the most important breakthroughs in modern cosmology is the discovery that the expansion of the Universe\footnote{We use capital letter ``Universe'' to refer to the universe we are actually living in, lower letter ``universe'' for a generic one.} is accelerating \cite{perlmutter, riess}. This suggests the existence of a positive cosmological constant, $\Lambda$, that could be a source for the vacuum energy density that drives said expansion. However, the field theoretical prediction for the vacuum energy density is some $10^{120}$ order of magnitudes larger than the observed value, so a considerable amount of effort has been put into explaining why the vacuum energy should in fact \emph{not} gravitate (see, e.g. \cite{1002.4275}), and something else has to account for the accelerated expansion of the Universe. 

One possibility is that the accelerated expansion could be driven by the inhomogeneity of the Universe, instead of some mysterious form of dark energy. Despite the assumption in the standard Friedmann-Lema\^{i}tre-Robertson-Walker (FLRW) cosmology that the Universe is homogeneous and isotropic, this could well be an over simplification, since there exist inhomogeneities on scale less than 150 Mpc \cite{1103.5974}. (See also, \cite{1211.6256}.) It is therefore conceivable that inhomogeneity might affect the expansion rate of the Universe  \cite{0605632, 0708.3622, 0711.3459, 0712.3457, 0801.0068, 1012.0784, 1106.1693, 1112.5335}. In fact, local inhomogeneities are important for several reasons. For example, \emph{even if} there is indeed a cosmological constant, effects from local inhomogeneities would result in an effective value, instead of the true value of the cosmological constant \cite{1104.0730}. In \cite{1311.1476}, it was pointed out that a local underdensity originated from a $3\sigma$  peaks of the primordial curvature perturbations field is sufficient to induce a correction to the value of a cosmological constant of the order of 1.5\%, which is not a small correction in the era of high precision cosmology. Furthermore, failure to take local inhomogeneities into account could lead to the wrong conclusion of an evolving dark energy, even though the Universe contains only a constant $\Lambda$ \cite{1006.4735}. In addition, there is an apparent discrepancy between local and large scale estimations of the current value of the Hubble parameter, which has been known for a few years now (see, e.g., \cite{1306.6766}). This tension has been made worse with recent observations. The recent result of Riess et al. \cite{riess2}, based on Cepheids, gives $H_0=73.24 \pm 1.74 ~\text{km} ~\text{s}^{-1} \text{Mpc}^{-1}$, which differs from the Planck's cosmic microwave background (CMB) result of $H_0=66.93 \pm 0.62 ~\text{km} ~\text{s}^{-1} \text{Mpc}^{-1}$ \cite{planck} by $3.4\sigma$. One possible way to resolve this tension is to consider local inhomogeneities, i.e., different probes are actually observing a different $H_0$ \cite{1609.04081, 1608.00534, 1706.09734}.

The most well-known inhomogeneous cosmology is probably the spherically symmetric Lema\^{i}tre-Tolman-Bondi (LTB) universes \cite{Le, Tolman, B}. Unlike FLRW universes in which both pressure $p$, and density $\rho$, of the fluid are assumed to be only functions of time, in LTB models the density can vary from place to place, i.e., $\rho=\rho(t,r)$. Nevertheless, pressure is still only a function of time. The properties of the LTB universes had been thoroughly investigated \cite{HL, HL1, HL2, HL3, 0709.2044}, and their observational prospects studied \cite{G}. 

Unfortunately, LTB cosmology suffers from quite a number of problems. First of all, there is a fine-tuning issue concerning our position in the center of the void: a mere 1\% displacement from the center would give rise to a significant dipole in the CMB \cite{0607334, 0909.4723}, beyond what has been actually observed. One could argue that perhaps by sheer coincidence we are at the center of the void, but this is no better than accepting the smallness of $\Lambda$ as being ``natural''.
In addition, the models fail to fit observations if various sources of data are fitted simultaneously \cite{1007.3725}. Specifically, the combined data sets of Type Ia supernovae (SNIa) and baryon acoustic oscillation (BAO) are in tension if one considers a wide enough redshift range for the BAO observations \cite{1007.3725, 1201.2790, 1408.1872, 1503.08045}. The tension further increases if Lyman $\alpha$ forest features of the BAO are included. In addition, taking into account simultaneously the supernovae observations, the local Hubble rate,  the small-angle CMB, and the kinematic Sunyaev-Zel'dovich (kSZ) effect, it has been shown that -- at least the simple versions of -- LTB models are effectively ruled out \cite{1108.2222}. These challenges to LTB models suggest that inhomogeneities of this type is not able to serve as an alternative to $\Lambda$, and at most only \emph{partially} account for the observed accelerated expansion. That is, one should still include $\Lambda$, which results in the so-called $\Lambda$LTB models. Even then, it has been argued that inclusion of $\Lambda$ does not enhance the viability of the models \cite{1601.05256}. 

There are, however, other inhomogeneous solutions to the Einstein field equations, that could perhaps achieve what LTB models have failed to do. In this work, we consider the Stephani universes (1967) \cite{Stephani, BF00759188}. It is ``complementary'' to the LTB models in the sense that it holds density $\rho$ constant in space, but allows pressure to vary, i.e., $\rho=\rho(t), p=p(t,r)$.  A Stephani universe is the most general conformally flat solution with an expanding perfect fluid source \cite{exact}.
The special case in which the model is spherically symmetric was known in the literature much earlier in the papers of Wyman (1946) \cite{wyman} and Kustaanheimo-Qvist (1948) \cite{KQ}. In this context, spherically symmetric means isotropic with respect to one point\footnote{Recall that if a universe is isotropic with respect to \emph{any} point then it is homogeneous. All our current observations were of course, effectively with respect to one point, namely our location in the Universe. Thus, observations have not ruled out an inhomogeneous Universe.}. 

A notable property of the Stephani universes is that the spatial scalar curvature is a function of time, $k=k(t)$. This means that spatial curvature could evolve and \emph{change sign} throughout the cosmic history. In contrast, the spatial scalar curvature in the LTB universes is a function of the spatial coordinates only. Both the Stephani and LTB models will reduce to FLRW universes in the homogeneous limit. Due to the time dependence in $k$, the global geometry and topology of Stephani universes are quite subtle, especially in the neighborhood when $k$ changes sign \cite{geom}. Though a generic, non-spherically symmetric, Stephani universe has no spacetime isometry, it admits three dimensional constant time hypersurfaces, which are maximally symmetric just like in the Friedmann models \cite{geom}. It is worth mentioning that the first inhomogeneous models that were tested with actual astronomical data were, in fact, the Stephani models, in the work of D\c{a}browski and Hendry \cite{Da}. On the other hand, the LTB models were first tested observationally a few years later by C\'el\'erier \cite{C} and Tomita \cite{Tomita}, and more recently by other authors \cite{biswas, 0909.4723, marra2011,valkenburg12}. Nevertheless, LTB models have been more extensively studied compared to the Stephani ones.

Recently, however, Balcerzak et al. performed a rather critical examination of some Stephani models and showed that they fall short of being a viable substitute for dark energy \cite{1409.1523}. In particular, using a joint constraint from SNIa BAO, CMB and the expected data from redshift drift, the authors showed that
unless the inhomogeneity parameter is sufficiently small  (thus rendering it ineffective to model the accelerated expansion), the behavior of the matter field in the current epoch deviates significantly from dust.  The authors remark that the Stephani models are therefore probably rather similar to the LTB ones, that is, one must still include $\Lambda$ in a realistic Stephani universe. This raises an interesting question: just how similar are these two classes of solutions to the Einstein Field Equations? After all, at least on first glance, their inhomogeneities are of very different nature. 


Observations aside, theoretical considerations are also important in putting constraints on a cosmological model.
While cosmology studies structures at the very large scales, one finds that, surprisingly, the search for the ultimate theory of quantum gravity -- which becomes important at the smallest scale, may have some implications to cosmology. It is well known that the Bekenstein-Hawking entropy of a black hole is proportional to the area of its event horizon. This suggests that the underlying degrees of freedom lives on the boundary of the black hole system, much like a hologram. Further investigations had led to the formulation of the ``holographic principle'' \cite{9310026, 9409089, 0203101}, which seems to be a robust feature of any consistent theory of quantum gravity. 
The holographic principle is, according to E. Witten, ``a real conceptual change in our thinking about gravity'' \cite{science}.
In string theory, we find the AdS/CFT correspondence \cite{maldacena}, which relates the gravitational physics in the bulk of an asymptotically Anti-de Sitter (AdS) spacetime with the physics of a quantum field theory on the boundary of the spacetime. This has allowed us to study, among other things, strongly coupled quantum system like the quark-gluon plasma produced at the RHIC collider, with toy models involving black holes in an asymptotically AdS spacetime \cite{1112.5403}.
The basic tenet of the holographic principle, even for spacetimes that are not asymptotically AdS, is this: the maximum number of degrees of freedom in a volume of space is bounded by the area of the boundary. That is, the holographic principle provides an \emph{entropy bound}. 

The applications of the holographic principle to cosmology was first carried out by Fischler and Susskind \cite{9806039}, and in the past decades had been extensively studied \cite{9811008, 9902173, 9902088, 9904120, 9904061, 9907012, 9908093, 9910185, 9911002, 9912055, 0006395, 0008140, 0404057, 0412218, 0501059, 0907.5542}. Such applications of the holographic principle are difficult tasks, primarily due to the absence of a natural boundary as in the case of the event horizon of a black hole, since most notions of ``horizons'' in cosmology are not globally defined \cite{1612.01084}. Nevertheless, a variety of applications in the cosmological context has been found, such as to deduce the most probable value of the cosmological constant \cite{0001145, 0403052}, to put a maximal bound on the number of $e$-foldings in inflation \cite{0307459,0308145,0312014} (see, however, \cite{0402162}), and to explain the suppression of the low multipoles in the CMB anisotropy power spectrum \cite{0501059, 0409275, 0412227}. 

In the work of Wang, Abdalla, and Osada \cite{0006395}, it was shown that in contrast to flat homogeneous models, the holographic principle may break down in some  LTB models. However, ``entropically realistic'' models (with model parameters chosen to reproduce the ``correct'' matter entropy of the current epoch; see also Footnote \ref{f7}) with fractal parabolic solutions always satisfy the holographic principle. This means that the holographic principle can serve as a test to rule out cosmological models, and/or to narrow down the allowed range of model parameters. However, satisfying the holographic principle is of course \emph{not} a guarantee that the models can fit all observational data. Even if one takes seriously the holographic principle as being fundamental, it is only a necessary condition, not a sufficient one, for a viable cosmological model. The LTB models provide just such an example -- a cosmology that satisfies the holographic principle but fails to fit observations. \emph{The main purpose of this paper is to study whether Stephani models behave in a similar way.} 

We found that this is indeed the case. Working with a specific model of the Stephani universe (the Stephani-D\c{a}browski model), we show that it is entropically viable -- both the holographic principle and the generalized second law are satisfied. However, from the observational point of view, the model is problematic. We show directly the tension that arises from the attempt to fit observational data to determine the model parameter. Specifically, if the value of the ``inhomogeneity parameter'' is chosen to fit $H_0$ (with some assumptions to be specified), then the model does not fit the supernovae constraint well. With some caveats, our finding supports the claim of Balcerzak et al. that Stephani models have problems fitting simultaneous constraints from observations. Their analysis, which concluded that the inhomogeneity parameter must be small is rather indirect, the argument presented is of the \emph{reductio ad absurdum} kind. Though less comprehensive, our simple analysis shows the difficulty of data fitting in a more direct manner. 

The remainder of this paper is organized as follows. In Sec.\ref{II} we will review the Stephani models, focusing on the spherically symmetric case. 
In Sec.\ref{III} we study the properties of the apparent horizon and show that the locations of the particle horizon and the apparent horizon in a specific Stephani model are not of the same order of magnitude, in contrast to the case of FLRW $k=0$ cosmology. In Sec.\ref{IV}, We calculate the entropy density and the total entropy value within the apparent horizon. Though the matter entropy decreases with time, the total entropy monotonically increases with time, in accordance to the (generalized) second law of thermodynamics. We also study the holographic principle in this model and found that again, much like in some LTB models, the holographic principle holds in this particular model of Stephani cosmology. We briefly comment on the energy conditions.
In Sec.\ref{V} we will discuss the observational aspects for the model. We show that, assuming the age of the Universe to be 13.7 billion years, the inhomogeneous parameter cannot simultaneously satisfy the observational data of Type Ia supernovae (using the Union2.1 data \cite{1105.3470}) and the Hubble parameter, whilst also serve to explain the accelerated expansion of the Universe, so the situation is much like that in the LTB models. 
In Sec.\ref{VI}, we summarize our results. The upshot is this: together with the previous study \cite{1409.1523}, our work suggests that -- at least the simple versions of -- Stephani models cannot by themselves hope to explain the accelerated expansion of the Universe, so at best one can only consider $\Lambda$-Stephani models, much like that for $\Lambda$-LTB.

\section{A Review of the Stephani Universes}\label{II}

The spherically symmetric Stephani models with a non-uniform pressure fluid is an exact solution of Einstein field equations that is conformally flat (Weyl tensor vanishes). The pressure in this model is non-uniform in the sense that it depends on both the temporal and spatial coordinates, namely $t$ and $r$. This is to be contrasted with the LTB model, in which the inhomogeneity arises from the non-uniform \emph{density} instead of the pressure. An interesting property of the Stephani universes is that comoving observers do not follow geodesics -- the expansion scalar $\Theta$, and the acceleration vector $\dot{u}^a$, are nonzero in this cosmology. This should be contrasted with the LTB universes, in which $\Theta$ and the shear $\sigma_{ab}$, are nonzero.

The metric tensor of a general Stephani universe is given by
\begin{equation}\label{0}
\mathrm{d}s^2=-c^2D^2 \mathrm{d}t^2 + \frac{R^2(t)}{V^2}(\mathrm{d}x^2+\mathrm{d}y^2+\mathrm{d}z^2),
\end{equation}
where
\begin{equation}
V=1+\frac{k(t)}{4}\left\{\left[x-x_0(t)\right]^2+\left[y-y_0(t)\right]^2 + \left[z-z_0(t)\right]^2\right\},
\end{equation}
\begin{equation}
D=F(t)\left(\frac{\dot{V}}{V}-\frac{\dot{R}}{R}\right) = F\left[\frac{\partial}{\partial t} \left(\frac{V}{R}\right)\right]\frac{R}{V},
\end{equation}
and
\begin{equation}
k=\left[C^2(t)-\frac{1}{c^2F^2(t)}\right]R^2(t),
\end{equation}
with $C,F,R, x_0,y_0,z_0$ being arbitrary functions of time.
The function $R(t)$ is the scale factor, measured in kilometers or
megaparsecs, while $F(t)$ is measured in seconds. The speed of light is 
$c=3\times10^5$ km/s. The spatial scalar curvature, which is a function of time, is dimensionless. The quantities
$D(r,t)$ and $V(r,t)$ are also dimensionless. 

This metric satisfies Einstein field equations with a perfect fluid source with energy density $\varepsilon$ satisfying
\begin{equation}
\frac{8\pi G}{c^4}\varepsilon = 3 C^2(t),
\end{equation}
and pressure $p$ satisfying
\begin{equation}
\frac{8\pi G}{c^4} p = -3 C^2(t) + 2 C(t) \left[\frac{\mathrm{d}C}{\mathrm{d}t}\right]\left[\frac{\partial}{\partial t}\left(\frac{V}{R}\right)\right]^{-1},
\end{equation}
where we have restored $G$ and $c$ explicitly for clarity.

From now onwards, we shall set $x_0=y_0=z_0=0$, so that the observer is situated at the center $r=0$.
Consequently,
\begin{equation}
V=1+\frac{1}{4}k(t) r^2,
\end{equation}
and so $\dot{V}=r^2\dot{k}/4$.

There are two exactly spherically symmetric Stephani models which will reduce to the Friedmann universe in the homogeneous limit \cite{dabrowski93}. The first model  satisfies the condition $(\partial^2/\partial t^2)({V}/{R})=0$ and the second one  fulfills the condition $(\d/\d t)\left({k}/{R}\right)=0$. In this paper, for simplicity we only investigate the second model which is characterized by an inhomogeneity parameter $\beta$. 
By imposing the condition $(\mathrm{d}/\mathrm{d}t)(k/R)=0$ and choosing the function $F$ such that\footnote{The sign is chosen so that \begin{equation}-\int \frac{F\dot{R}}{R} \mathrm{d}t = t\end{equation} in Eqs.(2-9) of \cite{0902.2899}.} $F\dot{R}/R=-1$, one can show that $D=1/V$. 
The metric then reduces to, with the speed of light still kept explicit,
\begin{equation}  \label{1}
\mathrm{d}s^2=-c^2 D^2(t,r)\mathrm{d}t^2+\frac{R^2(t)}{V^2(r,t)}\left[%
\mathrm{d}r^2+r^2\left(\mathrm{d}\theta^2+\sin^2 \theta\mathrm{d}%
\phi^2\right)\right].
\end{equation}
Furthermore, $C(t)=A\cdot R(t)$, where $A=\text{const.}$, and we will take the ansatz \cite{dabrowski93} 
\begin{equation}\label{MI}
\begin{cases}
R(t) = \beta t^2 + \gamma t + \eta, \\
V(r,t)=1-\frac{\beta}{c^2}(\beta t^2 + \gamma t + \eta) r^2,\\
k(t)=-\frac{4\beta R(t)}{c^2},\\
\gamma=\pm \sqrt{4\beta \eta + 1}.
\end{cases}
\end{equation}
Here $\beta$, $\gamma$, $\eta$ are parameters with dimensions  $[\beta]= \text{km}^2/(\text{s}^2\text{Mpc})$, $[\gamma]= \text{km/s}$ and $[\eta]=\text{Mpc}$, respectively\footnote{This ansatz is taken from \cite{dabrowski93}, and so should be more appropriately called the Stephani-D\c{a}browski model. That is, it is a very specific case of the Stephani class of solutions. In \cite{9911235}, Barrett and Clarkson referred to it simply as the D\c{a}browski model, which is referred to as ``Model I'' in \cite{0902.2899}; but not the same as models given in \cite{wesson}, which is marked in \cite{0902.2899} as Model II.}.

We note that the expansion scalar of this geometry is given by
\begin{equation}
\Theta = \frac{3(V\dot{R}-R\dot{V})}{R}.
\end{equation}
We can therefore define the Hubble parameter by
\begin{equation}
H:=\frac{\Theta}{3}=\frac{V\dot{R}-R\dot{V}}{R}.
\end{equation}
For $V=1+(1/4)kr^2$, with the imposed condition that $(\mathrm{d}/\mathrm{d}t)(k/R)=0$, we get simply the familiar looking
\begin{equation}
H=\frac{\dot{R}}{R}.
\end{equation}

By assuming a  perfect fluid  energy-momentum tensor $T^{\mu}_{~\nu}=(\rho c^2 +P)u^{\mu}u_{\nu}+P\delta^{\mu}_{~\nu}$, the time-time component of the Einstein equations, which is the generalized Friedmann equation, is given by
\begin{equation}\label{R}
\left(\frac{\dot{R}}{R}\right)^2=\frac{4\beta}{ R}+\frac{8\pi G}{3}\rho.
\end{equation}
Like the Friedmann models, the critical density is $\rho_{\text{cr}}={3H^2}/{8 \pi G}$. Define the density parameter as usual by $\Omega(t):={\rho(t)}/{\rho_\text{cr}(t)}$. By inserting the present value for the time, $t=t_0$, in Eq. (\ref{R}) we obtain
\begin{equation}
\frac{4\beta}{R_0 H^2_0}+\frac{8 \pi G}{3 H^2_0}\rho_0=
\Omega_{\text{inh},0}+\Omega_{m,0}=1,
\end{equation}
where $H_0$ is the present value of the Hubble parameter $H={\dot{R}}/{R}$, while $\Omega_{m,0}={\rho_0}/{\rho_{\text{cr},0}}$ is the matter density of the Universe, and $\Omega_{\text{inh},0}$ denotes the ``inhomogeneity density'', which is the dark energy density in this model. Explicitly, the inhomogeneity density is directly proportional to $\beta$:
\begin{equation}
\Omega_{\text{inh},0}=\frac{4\beta}{R_0 H^2_0}.
\end{equation}
Since density is positive, this imposes the condition that the inhomogeneity parameter $\beta$ should be positive in this subclass of models\footnote{Note that our sign of $\beta$ is opposite from that of \cite{1409.1523}.}. (Negative densities have nevertheless been considered in cosmological contexts in \cite{1402.4522}.) By the expression of $k(t)$ in Eq.(\ref{MI}), this also implies that $k(t) < 0$ at all time, so we are dealing with a negatively curved Stephani universe.

From the observational point of view, the inhomogeneous spherically symmetric Stephani models were first studied to provide a possible explanation for the observed accelerated expansion of the Universe \cite{stelmach01}.
It was shown that these models can fit the accelerating expansion of the Universe with the same accuracy as the $\Lambda$CDM model \cite{GS, Hashemi}. As it turned out, such models are characterized by a higher value of the density parameter $\Omega_{m,0}$ compared with the standard $\Lambda$CDM model. Furthermore, these models are consistent with the location of the CMB peaks. In \cite{1312.1567}, the exact luminosity distance and apparent magnitude formulas were applied to the Union2 557 SNIa sample \cite{1004.1711}
to constrain the position of a non-centrally located observer. It was found that even at $3\sigma$ confidence level, an observer outside the center $r=0$ in a spherically symmetric Stephani universe cannot be further than 4.4 Gpc away from it. This is comparable to the size of a void in LTB models. The authors also evaluated the best fit for the inhomogeneity density: $\Omega_{\text{inh}}=0.77$, for a specific model that allows a barotropic equation of state at the center of symmetry.

\section{Particle Horizon and Apparent Horizon in Stephani Cosmology}\label{III}

Recall that in a matter-dominated flat FLRW cosmology, the particle horizon size is given by
\begin{equation}
d_H (t)= a(t)r := a(t)\int_0^t \frac{c ~\text{d}t'}{a(t')}.
\end{equation}
In the convention in which the scale factor is unitless, 
\begin{equation}
d_H (t_0) = \int_0^{t_0} \frac{c}{a(t')} \text{d}t' = \int_0^{t_0}\frac{c}{\left(\frac{3H_0t'}{2}\right)^{2/3}}\text{d}t' \approx 10^{28} ~\text{cm} \approx 3240.78 ~\text{Mpc}.
\end{equation}
The size of the apparent horizon in the same universe is given by $c/H_0 \approx 4166.67 ~\text{Mpc}$. Note that these two horizons are comparable in size, of the order $O(10^4) ~\text{Mpc}$.

For the Stephani model, the particle horizon is
\begin{equation}
d_H (t_0) = \frac{R_0}{V_0}\cdot r_0, 
\end{equation}
where
\begin{equation}
r_0:=\int_0^{t_0} \frac{c}{R(t')} \text{d}t'.
\end{equation}
Explicitly in terms of the ansatz parameters,
\begin{equation}
d_H (t_0) = \frac{\beta t_0^2 + \sqrt{4\beta\eta+1}t_0 + \eta}{1-\frac{\beta}{c^2}(\beta t_0^2 + \sqrt{4\beta\eta+1}t_0+\eta)r_0^2}\cdot r_0, 
\end{equation}
with
\begin{equation}
r_0:=\int_0^{t_0} \frac{c}{\beta t^2 + \sqrt{4\beta\eta+1}t + \eta} \text{d}t' \approx 3.6052 \times 10^5.
\end{equation}

We must take extra caution at this point due to the nontrivial geometry -- in Stephani cosmology with $k<0$, the spatial section actually consists of \emph{two disconnected sheets} \cite{geom}. To see this, consider the spatial distance \cite{geom}
\begin{equation}
\ell(t_0):=\int_0^{r_0} \frac{R(t_0)}{V(t_0,r)}\text{d}r = \frac{R(t_0)}{|k(t_0)|^{1/2}}\ln\left[\frac{1+\frac{1}{2}|k(t_0)|^{1/2}r_0}{|1-\frac{1}{2}|k(t_0)|^{1/2}r_0|}\right].
\end{equation}

It is clear that $\ell(t_0)$ diverges as one approaches the ``branch point''
\begin{equation}\label{branchpoint}
r_B(t_0) := \frac{2}{|k(t_0)|^{1/2}}.
\end{equation}
The spatial section $t=t_0$ thus consists of two disjoint sheets: the ``near sheet'' $r < r_B(t_0)$, and the ``far sheet'' $r > r_B(t_0)$. This is true for all spatial slices of fixed $t$. The far sheet is of no physical interest since the observer is situated at $r=0$, which is in the near sheet. For our model,
\begin{equation}
k = -\frac{4\beta R_0}{c^2} \approx -2.8573 \times 10^{-11}, 
\end{equation}
and 
\begin{equation}
r_B(t_0) = 3.7416 \times 10^5.
\end{equation}
The fact that $k$ is extremely small is, by itself, a good thing, since the observed spatial curvature of the Universe is close to being flat.

The particle horizon corresponds to $r_0 \approx 3.6052 \times 10^5$, which is rather close to the branch point, but nevertheless is still within the near sheet, and is thus physical. Evaluating $d_H(t_0)$, we obtain $4.0779\times 10^{29} \text{cm}$, which is one order of magnitude larger than the value for flat FLRW universe.

For a spherically symmetric spacetime, the dynamical apparent horizon can be obtained from the condition $|\nabla r|^2:= \langle \nabla r, \nabla r\rangle=0$, with the result \cite{wald, Hawking}
\begin{equation}\label{AH}
r_{\text{AH}}^2 \dot{R}^2(t) + \frac{k(t)}{2}r_{\text{AH}}^2 - \frac{k^2(t)}{16}r_{\text{AH}}^4 -1 =0,
\end{equation}
that is,
\begin{equation}
r_{\text{AH}}^2 (2\beta t + \sqrt{4\beta\eta+1})^2 - \frac{2\beta(\beta t^2 + \sqrt{4\beta\eta+1}t+\eta)}{c^2}r_{\text{AH}}^2-\frac{\beta^2(\beta t^2 + \sqrt{4\beta\eta+1}t+\eta)^2}{c^4}r_{\text{AH}}^4-1=0.
\end{equation}
If we substitute in the values of the various parameters, we would obtain two positive solutions\footnote{The large difference between the values of the two roots is due to large numbers involved in the coefficients, such as $c$. If we have substituted in $\beta=\eta=t=c=1$ instead, we would get two positive roots of the same order of magnitude:
\begin{equation}
r_1 = \frac{1}{2}\left(3-\sqrt{5}\right), ~~r_2=\frac{1}{2}\left(-1+\sqrt{5}\right).
\end{equation}
We note a remarkable fact (most likely a coincidence) that $r_2$ is exactly the golden section conjugate.}:
\begin{equation}
r_1 \approx 0.5291, ~~r_2 \approx 2.6456 \times 10^{11}.
\end{equation}

Since the root $r_2$ is located in the far sheet, let us consider only $r_1$. The \emph{physical} apparent horizon is
\begin{equation}\label{AHphy}
\tilde{r}:=r\frac{R(t)}{V(r,t)}=\frac{c}{\sqrt{\left(\frac{\dot{R}(t)}{R(t)}\right)^2+\frac{k(t)}{R^2(t)}}}.
\end{equation}
We note that if $k(t)=0, \pm 1$, then Eq.(\ref{AHphy}) will reduce to the physical apparent horizon for the FLRW universes \cite{9902173}, where it is just the Hubble radius in the case of a spatially flat Friedmann solution. 

With the aforementioned value of $r_1$, Eq.(\ref{AHphy}) gives at present time, 
\begin{equation}
\tilde{r}_1 \approx 0.0139 ~\text{Mpc} \sim 10^{22} ~\text{cm}.
\end{equation}
This value is much smaller than the particle horizon $10^{29} ~\text{cm} \approx  32407.7929~\text {Mpc}$. 
Therefore, unlike the flat FLRW universe case, in which the size of the particle horizon is comparable to that of the apparent horizon, the apparent horizon in this Stephani model is much smaller than its particle horizon.

\section{Entropy and Holographic Principle}\label{IV}

In this work, we will take the holographic principle to mean the inequality $S \leqslant A/(4l_p^2)$, where $l_p$ is the Planck length. 
To apply the holographic principle in the Stephani models, we first need to identify a boundary surface. Unlike the case of a black hole, there is no natural, globally defined boundary, like the event horizon. Following \cite{0006395}, we shall use the apparent horizon for this purpose. Although an apparent horizon depends on the choice of spacetime foliation, it is a natural surface in the context of cosmology. After all, foliation dependence is less of a concern here -- the fact that the Universe looks more or less homogeneous and isotropic, and that its spatial curvature looks flat, are all statements that are based on a special choice of frame. Furthermore, since apparent horizon is defined quasi-locally it always exists in most cosmological contexts, unlike say, the particle horizon.  In addition, the first and second law of thermodynamics seem to hold when one works with apparent horizon in the context of an accelerated expanding (homogeneous and isotropic) universe driven by dark energy of time-dependent equation of state, but not the cosmic event horizon \cite{0511051}.
Working with the apparent horizon is therefore well motivated \cite{9902173, 1106.4427}. 

In order to calculate the entropy (of the matter field) within the apparent horizon, we need to specify the local entropy density. As is well-known, when a particle species becomes non-relativistic and decouples from the primordial plasma of the early Universe, its entropy will be transferred to other particles which are still in thermal equilibrium with the plasma (see any cosmology textbook, e.g., \cite{kolb}). Consequently, the massless particles such as photons dominate the entropy of the Universe\footnote{We emphasize that here we only discuss the matter degrees of freedom. The entropy for matter alone can in fact decrease in an expanding universe \cite{1306.2714}.
The total entropy budget of any realistic universe should of course include gravitational entropy, and entropy as a whole is increasing (the second law of thermodynamics), which resuts in the observed arrow of time. 
}.
It is therefore fair to assume that the entropy of the Universe is produced before the dust-filled era (photons produced by stars are negligible).

From thermodynamics, the local entropy density for the dust-filled Universe\footnote{For a spherically symmetric Stephani cosmology, the matter is a fluid with nonzero pressure gradient. The dust assumption is only valid for an observer located at $r=0$.}, still dominated by relativistic particles is, $s=(\epsilon+P)/T$. Here $\epsilon=\rho c^2$ is
the radiation energy density given by $\alpha T^{4}$, with $T$ being the
temperature of the Universe, $\alpha$ being the radiation constant, and $P=\rho c^2$/3 is the pressure of the radiation. 
Therefore, $s=(\epsilon+P)/{T} = (4/3)(\rho c^2/T)$. 
Assuming that during the expansion of the Universe, the radiation is that of a blackbody, and by assuming that the number density of photon is (approximately) conserved, for the inhomogeneous background we have the following relations
\begin{equation}  \label{10}
\frac{\hbar \nu'}{k_\text{B}T'}=\frac{\hbar\nu}{k_\text{B} T},\quad \nu'=\nu (1+z),
\end{equation}
where $\hbar$ is the reduced Planck constant, and $k_\text{B}$ is the Boltzmann constant. 

The general expression for redshift in any cosmology is given by \cite{EJ,eliis}:
\begin{equation}  \label{11}
1+z=\frac{\nu_\text{e}}{\nu _\text{o}}=\frac{(u_al^a)_\text{e}}{(u_bl^b)_\text{o}},
\end{equation}
where the indices \lq\lq{}o\rq\rq{}  and \lq\lq{}e\rq\rq{}  stand for the observer  and the
emitter positions, respectively. Here $l^a={\mathrm{d}x^a}/{\mathrm{d}s}$ is
the vector tangent to the null geodesics, $x^a=(t,r,\theta,\phi)$, with $s$ being an affine
parameter, and $u^a$ is the corresponding 4-vector velocity whose only nonvanishing
component being $u_{t}=-cD(r,t)$.




For metric (\ref{1}), one obtains \cite{0902.2899}
\begin{eqnarray}
l^{t} & = & \frac{V^2}{FR^2} \left[\left( \frac{V}{R} \right)_{,t}\right]^{-1} ,
\\
l^{r} & = & \pm \frac{V^2}{R^2} \sqrt{ 1 - \frac{h^2}{r^2}},
\end{eqnarray}
as well as $l^{\theta} = 0$ and $l^{\varphi} = h{V^2}/{R^2r^2}$.

Therefore we have
\begin{equation}
l^t u_{t}=-c\frac{V(r,t)}{R(t)},
\end{equation}
and consequently, the expression for the redshift, Eq. (\ref{11}), becomes
\begin{equation}
1+z=\frac{V_\text{e} R}{R_\text{e}V}.
\end{equation}
From Eq. (\ref{10}) we have $
T'=T(1+z)$, thus the local entropy density is given by
\begin{equation}
s(r,t)=\frac{\rho c^2 }{T}=\frac{4}{3}\frac{\alpha T^{4}}{T}=\frac{4}{3}\alpha T^{3}=\frac{4}{3}\frac{\alpha
T'^{3}}{(1+z)^{3}}.
\end{equation}%
Finally, the total entropy inside the apparent horizon is obtained via
\begin{equation}\label{S}
S_m=\int_{0}^{r_\text{AH}}s(t,r)~\mathrm{d}v,
\end{equation}%
where $\mathrm{d}v$ is the volume element of metric (\ref{1}), namely, $\mathrm{d}v=R^{3}(t)(V^{-3}(r,t))r^{2}\sin {\theta }~\mathrm{d}r\mathrm{d}\theta \mathrm{d}\phi $.
We have emphasized with a subscript $m$ that this entropy is the entropy of the matter sector only.

With the above equations, the total entropy, Eq. (\ref{S}), can be calculated: 
\begin{equation}\label{30}
S_m = \frac{16\pi}{3}\alpha T'^{3}\left(\frac{R(t)}{V(r,t)}\right)^3_\text{e}\int_{0}^{r_\text{AH}}r^{2}\mathrm{d}r=\frac{16\pi}{9}\alpha T'^{3}r_\text{AH}^3\left(\frac{R(t)}{V(r,t)}\right)^3_\text{e}.
\end{equation}
In Eq. (\ref{30}), the emitter position is
\begin{equation}
r_\text{e} = \int_0^{t_\text{e}} \frac{c}{R(t)} \d t,
\end{equation}
and the radiation constant is 
\begin{equation}
\alpha=\frac{\pi^2 k_B^4}{15 c^3 \hbar^3}.
\end{equation}
Taking $t_\text{e} \sim 10^{11} \text{s}$, the time at the end of radiation era, and the associated temperature $T_\text{e} \sim 1~\text{eV} \approx 11600~\text{K}$, we find that the apparent horizon is located at 
\begin{equation}
r_\text{AH} = r_1(t_\text{e}) \approx 0.8305.
\end{equation}
Evaluating the entropy in Eq.(\ref{30}), we obtain
\begin{equation}
S_m\approx 1.0742 \times 10^{58}~ \text{J/K}. 
\end{equation}
In natural units, this is $S_m \sim 10^{81}$. 

\begin{figure}
\centering
\includegraphics[width=5.0in]{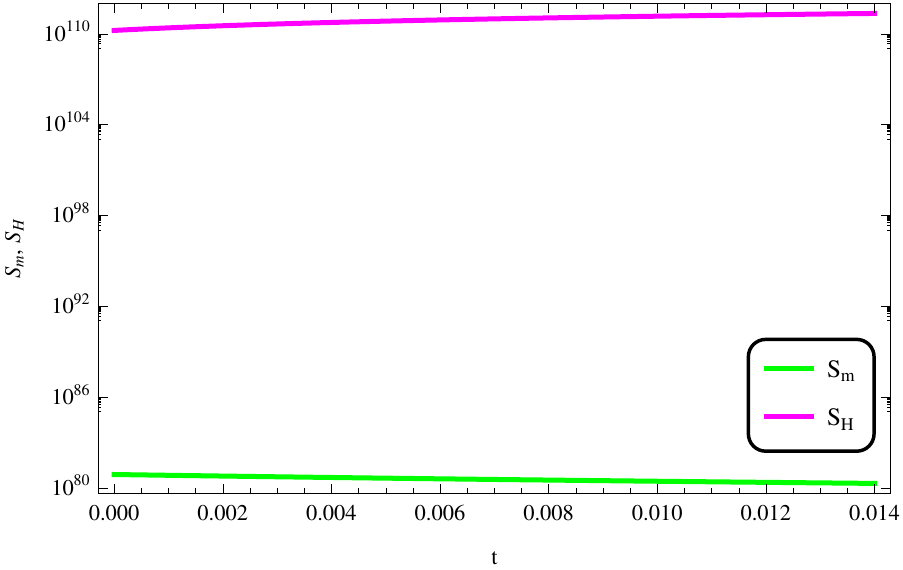}
\caption{The plots for the matter entropy $S_m$ (bottom), and the apparent horizon entropy $S_H$ (top) against time. The entropies are in natural units (hence dimensionless), while time is measured in the unit sMpc/km. Note that the holographic principle requires that $S_m < S_H$. This is always satisfied in this model. The sum $S_H + S_m$ always satisfies the generalized second law, i.e., it increases with time.}
\label{fig2}
\end{figure}

It is usually said that the entropy of the present observable universe is of the order $10^{90}$. That value is of course based on the usual assumption of flat FLRW cosmology. The fact that we obtained a much smaller number is not a cause for alarm since we have used the apparent horizon, which in this model is far smaller than the particle horizon. In addition, the value of the matter entropy is model-dependent and by itself does not invalidate a model unless the value is extremely small or extremely large\footnote{In \cite{0006395}, the authors use $S \sim 10^{90}$ as an input to constrain the LTB models. In this work, we already have the model, its parameters fixed by observational requirements. Our input is instead, the age of the Universe.\label{f7}}. In fact, if one were to calculate the entropy within the apparent horizon in the present epoch, one would find a \emph{smaller} value, $S_m \sim 10^{80}$. In this cosmology, $r_1[t_\text{e}] > r_1[t_0]$. In fact, the coordinate radius of the apparent horizon is decreasing with time, and so does the matter entropy. However, this is not a violation of the second law of thermodynamics, since although $r_1$ decreases as a function of time, the \emph{physical area}
\begin{equation}
A=\frac{4\pi r_1^2 R^2(t)}{V^2(r,t)}
\end{equation}
is increasing in time. See Fig.(\ref{fig2}). At $t=t_\text{e}$, the entropy associated with the apparent horizon is $S_H \sim 10^{110}$ in natural units. This increases to $S_H \sim 10^{111}$ at the current epoch. Since $S_H$ is much larger than the matter entropy, their sum is dominated by $S_{H}$, so it is clear that the sum of the two entropies increases with time, and so the (generalized) second law is not violated. In addition, the holographic principle is also respected by a large margin. 

We emphasize again that the decrease in the matter entropy as the Universe expands is not without precedence -- a similar phonemenon has been observed even in a flat FLRW universe with a positive cosmological constant \cite{1306.2714}. However, there are some differences: firstly, the cosmological event horizon was used in \cite{1306.2714}, not the apparent horizon, but most importantly, whereas $S_m$ is monotonically decreasing in the Stephani model we study, it exhibits an increasing phase at early times in the flat FLRW case when the Universe is decelerating. 

Finally we comment on the energy conditions. A perfect fluid energy-momentum tensor given by $T^{\mu}_{~~\nu}= (\rho c^2+P)u^{\mu}u_{\nu}+P \delta^{\mu}_{\nu}$, with energy density given by
\begin{equation}
\rho= \,{\frac {3}{8\pi\,G \left( \beta\,{t}^{2
}+\gamma\,t+\eta \right) ^{2}}}>0.
\end{equation}

Let us consider the following three inequalities:
\begin{equation}\label{e1}
\rho+3P/c^2=-\,{\frac {\,{3\beta}{r}^{2}/c^2}{4\pi\,G \left( \beta
\,{t}^{2}+\gamma t+\eta \right)}} > 0,
\end{equation}
\begin{equation}\label{e2}
\rho+P/c^2= \,{\frac {1-\,{\beta}{r}^{2}(\beta\,{t}^{2
}+\gamma\,t+\eta)/c^2}{4\pi\,G \left( \beta
\,{t}^{2}+\gamma t+\eta \right) ^{2}}} > 0,
\end{equation}
and
\begin{equation}\label{e3}
\rho-P/c^2=\,{\frac {2+\,{\beta}{r}^{2}(\beta\,{t}^{2
}+\gamma\,t+\eta)/c^2}{4\pi\,G \left( \beta\,{
t}^{2}+\gamma t+\eta \right) ^{2}}} > 0,
\end{equation}
If Eq.(\ref{e1}) is satisfied, then the strong energy condition (SEC) is satisfied. Similarly, the weak energy condition (WEC) is satisfied if Eq.(\ref{e2}) holds. The dominant energy condition (DEC) is satisfied if both Eq.(\ref{e2}) and Eq.(\ref{e3}) are satisfied. These energy conditions were previously investigated in \cite{0410033} in the context of future sudden singularities.
We see that SEC cannot be satisfied since $R(t)= \beta t^2 + \gamma t + \eta > 0$. However, the WEC is satisfied if and only if
\begin{equation}\label{wec}
r^2(\beta t^2 + \gamma t + \eta)=r^2R(t) <\frac{c^2}{\beta}.
\end{equation}
Since Eq.(\ref{e3}) always holds, DEC holds whenever WEC does. The reason we want to discuss the energy conditions is as follows: the generalized second law of thermodynamics is a consequence of null energy condition (NEC) \cite{davies, 1405.0403}. Our claim that the generalized second law holds (and also that physical apparent horizon size is increasing) should be consistent with this. NEC is the weakest of the energy conditions -- it holds as long as WEC does. Interestingly, we note that although WEC is violated for sufficiently large coordinate radius $r$, it is \emph{always} satisfied in the ``near sheet'', where $r < r_B$, with $r_B$ defined in Eq.(\ref{branchpoint}), and $k = -4\beta R/c^2$. In fact Eq.(\ref{wec}) is \emph{exactly} the condition that $r$ should be within the ``near sheet''. Therefore DEC, WEC, and NEC hold in the Stephani-D\c{a}browski model since we are confined to the ``near sheet'' of the geometry. This is consistent with our findings that apparent horizon is increasing in size, and that the generalized second law holds.

\section{The Challenge from Observational Constraints}\label{V}

Let us take $H_0=72 ~\text{km} ~\text{s}^{-1}\text{Mpc}^{-1}$. (The conclusion is not very sensitive to the exact choice of $H_0$, we can relax it in the range $60 \leqslant H_0 \leqslant 80$.) Note that
\begin{equation}
H_0 = \frac{2\beta t_0 + \gamma}{R_0} = \frac{2\beta t_0 + \gamma}{4\beta} H_0^2 \Omega_{\text{inh},0}.
\end{equation}
This yields 
\begin{equation}\label{gamma}
\gamma = 2\beta\left[\frac{2}{H_0\Omega_{\text{inh,0}}}-t_0\right],
\end{equation}
which provides a constraint on the sign of the parameter $\gamma$. Namely, $\gamma > 0$ if $t_0 < {2}/(H_0 \Omega_{\text{inh,0}}) = {R_0 H_0}/(2\beta)$.
Taking $\Omega_{\text{inh,0}}=0.72$, we get  ${2}/(H_0 \Omega_{\text{inh,0}})\approx 0.03858$. Taking the age of the Universe as 13.7 billion years, or $t_0=0.014 ~\text{sMpc}/\text{km}$, we see that $\gamma >0$, so we take $\gamma=\sqrt{4\beta\eta+1}$. Note that if we allow a range $60 \leqslant H_0 \leqslant 80$, and $0.6 \leqslant \Omega_{\text{inh,0}} \leqslant 0.8$, then $0.03125 \lesssim {2}/(H_0 \Omega_{\text{inh,0}}) \lesssim 0.05556$. Even the lower end of this range is still larger than $t_0$. Of course the age of the Universe in this Stephani model can be tuned by choosing different values of the various parameters, the inequality can be satisfied as long as the age difference is not too large. We choose the commonly accepted age of the Universe, 13.7 billion years, as the input in this work. 

Thus, equating $\gamma=\sqrt{4\beta\eta+1}$ with Eq.(\ref{gamma}), and with the value of $t_0$ substituted, we have
\begin{equation}\label{sq1}
\sqrt{4\beta\eta+1}\approx 0.04916\beta.
\end{equation}
On the other hand, $H_0$ gives
\begin{equation}\label{sq2}
72 = \frac{2\beta t_0 + \sqrt{4\beta\eta+1}}{\beta t_0^2 + \sqrt{4\beta\eta+1}+\eta}.
\end{equation}
The simultaneous equations Eq.(\ref{sq1}) and Eq.(\ref{sq2}) can be solved to give the value of $\beta$ and $\eta$. We have:
\begin{equation}
\beta \approx 24.4925, ~~\eta \approx 0.004591.
\end{equation}

We can now test whether this model satisfies other observations, such as SNIa. There is, however, a caveat worth emphasizing: due to the varying pressure, a test particle that follows a geodesic will not be comoving in this model. Even if a cluster of objects are initially comoving, it is not clear if they will remain comoving as the Universe expands \cite{1409.1523}. The redshift in standard cosmology $\lambda_0/\lambda_\text{e}=a_0/a_\text{e}$, where ``e'' denotes quantities associated with the emitter position, is modified to be
\begin{equation}
\frac{\lambda_0}{\lambda_\text{e}} = \frac{a_0}{a_\text{e}} V_\text{e}
\end{equation}
in this Stephani cosmology. As emphasized in \cite{1409.1523}, this is true only for light emitted by comoving sources. It is therefore crucial to know if the astrophysical sources that we are considering are, in fact, comoving. In \cite{1409.1523}, it is assumed that the departure from a comoving motion is sufficiently small so that we may treat SNIa as comoving objects. We will make the same assumption here.

The apparent magnitude $m$ is given by the magnitude-redshift relation \cite{KS, eliis}
\begin{equation}
\label{MBOL}
m = M - 5\log_{10}{ \left( u_{a;b}L^{a}L^{b} \right)_\text{o}} + 5\log_{10}{cz}
+ \frac{5}{2} \left( \log_{10}{e} \right) \left\{ \left( 4 - \frac {u_{a;bc}L^{a}
L^{b}L^{c}}{ \left( u_{a;b}L^{a}L^{b} \right)^2} \right) z +
\mathcal{O} \left( z^2 \right) \right\}_\text{o},
\end{equation}
where  $M$  is the absolute magnitude, $ u_{a;b}  =  ({1}/{3}) \Theta h_{ab} - \dot{u}_{a} u_{b}$, $ L^{a}  := {l^{a}}/(u_{b}l^{b})$, $  h_{ab}  :=  g_{ab} + u_{a}u_{b}$, and $ u_{a}u^{a} =  - 1$. Here the semicolon in the subscript denotes a covariant derivative as usual. The quantity $\Theta$  is the expansion scalar,  while $\dot{u}_{a}$ is the acceleration vector, whereas $h_{ab}$ is the operator that projects vectors onto spacelike hypersurfaces. 

Following the same procedure as performed in \cite{0902.2899}, we can obtain 
\begin{equation}\label{mM}
m=M+25+5\log_{10}\left[cz\left(\frac{\beta t^2-\gamma t+\eta}{2\beta t-\gamma}\right)\right] +1.086z\left[1+2\beta \frac{(\beta t^2-\gamma t+\eta)}{(2\beta t-\gamma)^2}\right].
\end{equation}
Note that the luminosity distance is \cite{1409.1523}:
\begin{equation}
d_L(z) = \frac{2c(1+z)}{H_0} \sqrt{\frac{x^{-1}-(1+z(x))}{\Omega_\text{inh,0}}},
\end{equation}
where $x:=R/R_0$ is defined implicitly via the relation
\begin{equation}
z(x) = x^{-1}-1- \frac{ \Omega_{{\rm inh}, 0}}{4}  \left(\int ^1_x \frac{{\rm d}x\rq{}}{\sqrt{ x\rq{}^{-1}\Omega_{\rm inh,0}+x\rq{}^{-3(1+\omega)} (1-\Omega_{\rm inh,0})}}\right)^2.
\end{equation}
The distance modulus is thus
\begin{equation}\label{mu}
\mu(z) = 5 \log_{10} \left( 2c \frac{(1+z)}{H_{0}}\sqrt{\frac{x^{-1}(z)  -(1+z)}{\Omega_{{\rm inh, 0}}}}\right) +25,
\end{equation}
which leads to Eq.(\ref{mM}) above. Note that Eq.(\ref{mM}) is of course an approximation that neglects redshift of $\mathcal{O}(z^2)$ and above, however it is sufficiently good for observations fitting, since SNIa data are for low redshift $z \in (0,2)$.

\begin{figure}
\centering
\includegraphics[width=5.0in]{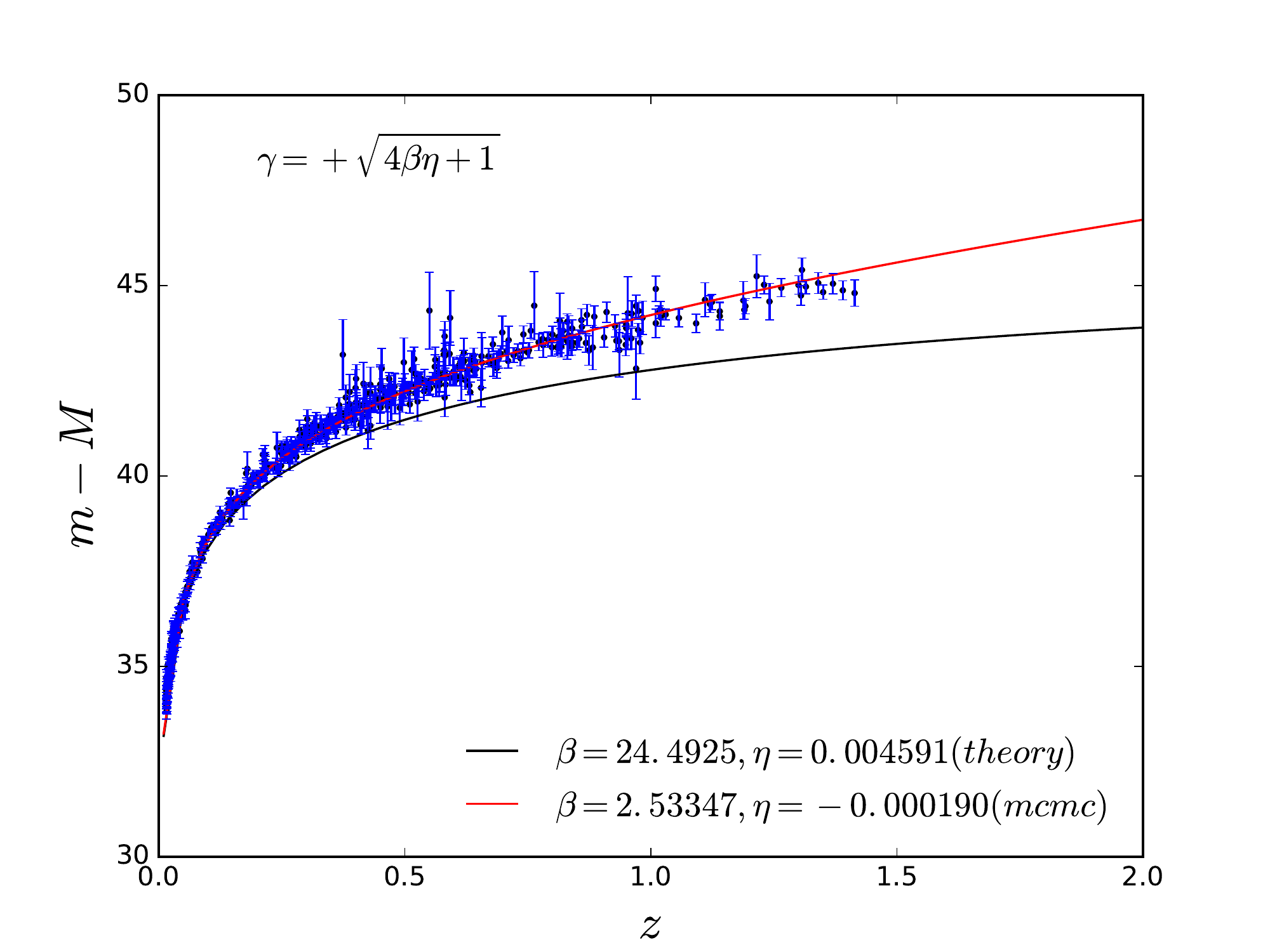}
\caption{The apparent magnitude $m-M$ with respect to the redshift $z$. The bottom curve (black) corresponds to the Stephani model with parameters $\beta=24.4925$ and $\eta=0.004591$, which are fixed to satisfy dark energy density and the Hubble parameter simultaneously. The top curve (red) corresponds to the best fit obtained via MCMC method.}
\label{fig1}
\end{figure}

The plot of Eq.(\ref{MBOL}), utilizing the SNIa data from Union2.1, is shown in Fig.(\ref{fig1}). The model with values of $\beta=24.4925$ and $\eta=0.004591$ obtained from considering the Hubble parameter and the dark energy density (labeled ``theory''), does not fit the data well. The best fit from utilizing Markov Chain Monte Carlo (MCMC) method yields instead $\beta = 2.53347_{-0.68105}^{+0.68621}$, and $\eta  = -0.000190_{-0.000115}^{+0.000116}$. Thus MCMC prefers the inhomogeneity parameter $\beta$ to be an order of magnitude smaller, and in addition $\eta$ should be of a different sign and smaller in magnitude.

This shows that the model cannot simultaneously satisfy constraints from SNIa, the Hubble parameter, and the dark energy density. 

\begin{figure}
\centering
\includegraphics[width=5.0in]{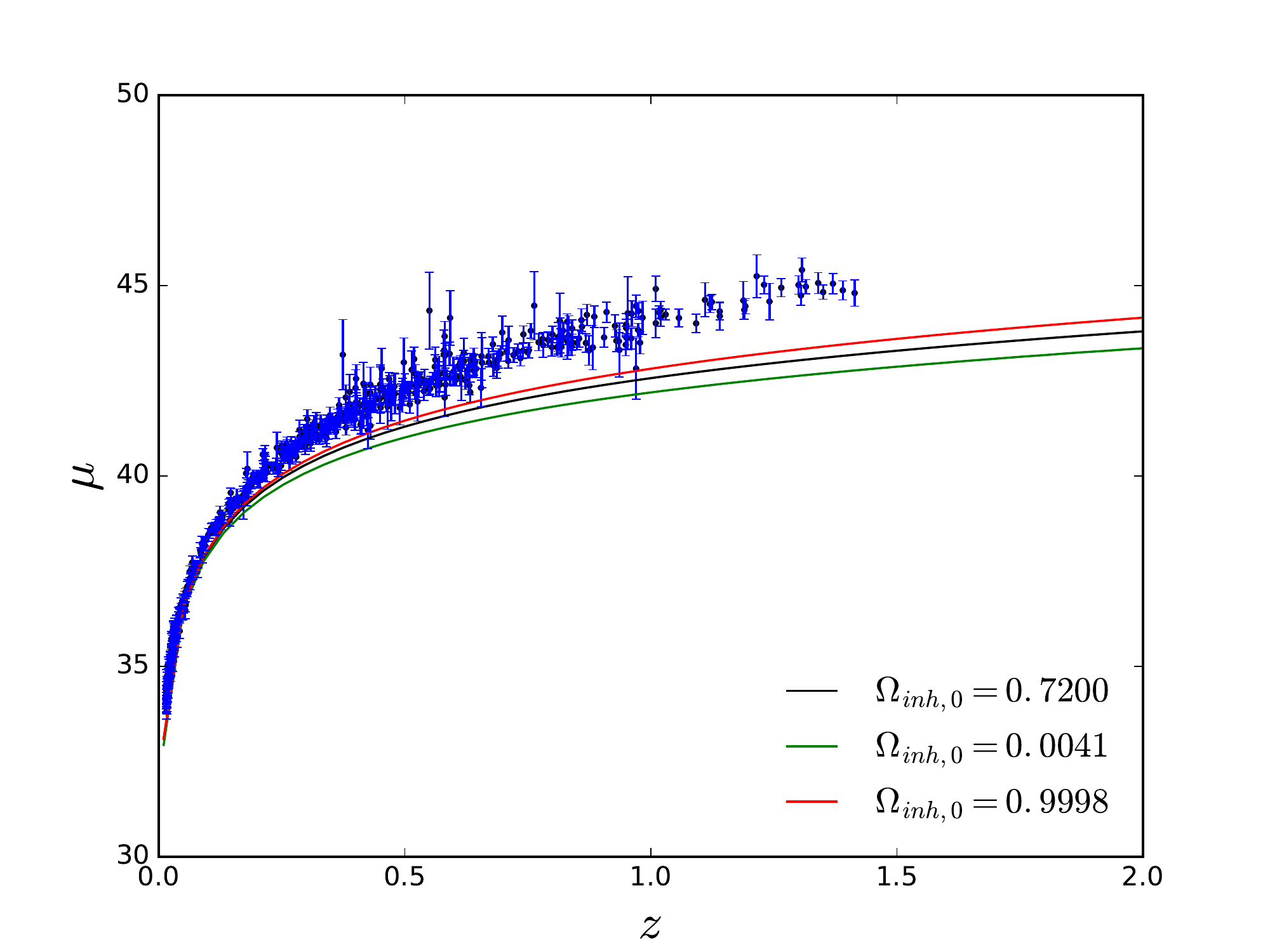}
\caption{The apparent magnitude $\mu$ with respect to the redshift $z$, with various of values $\Omega_{\text{inh},0}$. The top curve (red) corresponds to   $\Omega_{\text{inh},0}=0.9998$, the middle curve (black) corresponds to  $\Omega_{\text{inh},0}=0.7200$, and  the bottom curve (green) corresponds to $\Omega_{\text{inh},0}=0.0041$.
Since $\Omega_{\text{inh},0}$ can only take values between 0 and 1, this means the model cannot fit observational data.} 
\label{SNIa}
\end{figure}

To be more confident of our results, we can consider the full expression Eq.(\ref{mu}) instead of the approximation Eq.(\ref{MBOL}).
Furthermore, we observe that instead of fitting the model parameters $\eta$ and $\beta$, we can instead varies $\Omega_{\text{inh},0}$. This is because they are related via Eq.(\ref{gamma}), and $\Omega_{\text{inh},0}$ takes values between 0 and 1. By doing so we obtain Fig.(\ref{SNIa}). We see that in the admissible range of $\Omega_{\text{inh},0}$, we cannot fit supernovae data. (We show 3 representative values in the plot: a small value $0.0041$, corresponding to a matter-dominated universe, a large value $0.9998$ which corresponds to a universe that is almost all dark energy, and 0.72, which is close to the observed value.)
That is, the best fit $\beta$ and $\eta$ from MCMC above cannot actually satisfy $0<\Omega_{\text{inh},0} < 1$.

\begin{figure}
\centering
\includegraphics[width=5.0in]{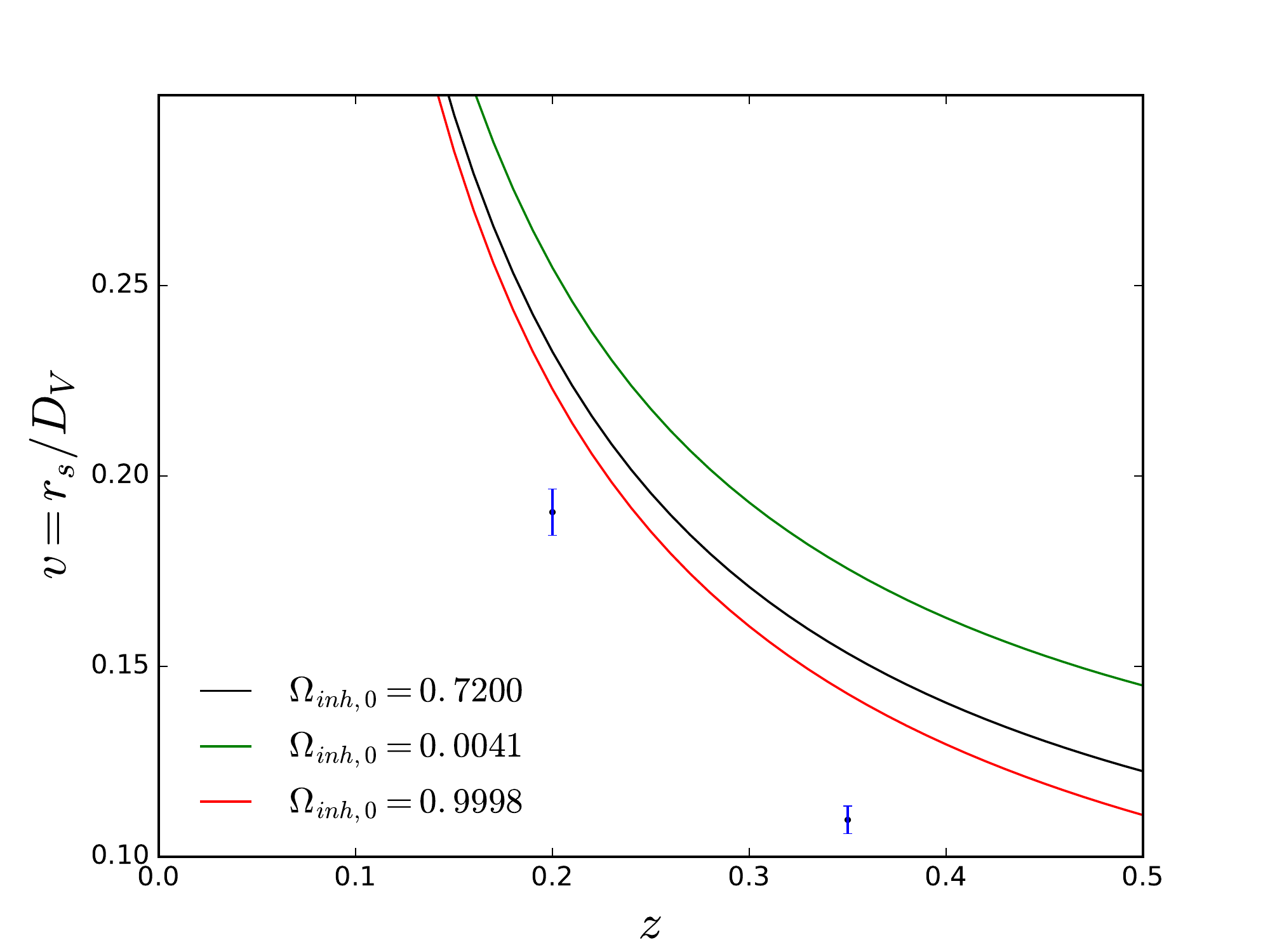}
\caption{$v=r_s/D_V$ with respect to the redshift $z$, with various of values $\Omega_{\text{inh},0}$. 
The top curve (red) corresponds to   $\Omega_{\text{inh},0}=0.9998$, the middle curve (black) corresponds to  $\Omega_{\text{inh},0}=0.7200$, and  the bottom curve (green) corresponds to $\Omega_{\text{inh},0}=0.0041$. Since $\Omega_{\text{inh},0}$ can only take values between 0 and 1, this means the model cannot fit observational data.} 
\label{BAO}
\end{figure}

Since sole supernovae data is often not enough to constrain observationally the parameters of the cosmological models \cite{1312.1567, 1409.1523}, let us be more comprehensive in the analysis. 
We can consider the volume distance \cite{1409.1523}
\begin{equation}\label{dv}
D_{{\rm V}}(z) = \frac{c}{H_{0}} \left[\frac{4z}{\Omega_{{\rm inh},0}}\frac{x^{-1}(z)-(1+z)}{h(x(z))} \right]^{1/3},
\end{equation}
where $h(x) = H(x)/H_{0}$, to investigate how well the model can fit data from baryon acoustic oscillations (BAO). The result is shown in Fig.(\ref{BAO}): where we have plotted the quantity 
\begin{equation}
v=\frac{r_s}{D_V},
\end{equation}
against redshift $z$. Here $r_s$ denotes the size of the comoving sound horizon during the baryon dragging epoch. See \cite{1409.1523} for details. For the admissible range of $\Omega_{\text{inh},0}$, all the curves obtained deviate quite far away from the two observational data (here to allow comparison with the results in \cite{1409.1523}, which employed the BAO data at $z=0.2$ and $z=0.35$, we also do the same).

The shift parameter, which is used in the context of CMB ($z=1090$) is given by
\begin{equation}
 \mathcal{R} =  \frac{H_{0}}{c}\sqrt{1-\Omega_{{\rm inh,0}}} (1+z_{d})^{-1}d_{L}=2 \sqrt{1-\Omega_{{\rm inh,0}}} \sqrt{\frac{(\frac{1}{x_{d}} - (1+z_{d}))}{\Omega_{{\rm inh}, 0}}},
\end{equation}
where $r_d \equiv r(x(z_d))$ is the coordinate distance at decoupling (again, see \cite{1409.1523}).
This quantity was estimated from 7-year WMAP observations \cite{1001.4538} to be
\begin{equation}
\mathcal{R}=1.725 \pm 0.018.
\end{equation}
For the model under study, we found that for $\Omega_{\text{inh},0} = 0.72, 0.0041, 0.9998$, we get $\mathcal{R}=0.2683, 0.8468, 0.0086$, respectively. Thus, again we see that this model does not fit data. (In \cite{1409.1523}, SNIa and BAO data fit well if one allows the fluid to have significant departure from dust, but even then the CMB shift constraint cannot be simultaneoulsy satisfied.)

From the above analysis of supernovae and BAO data, as well as the shift parameter, we see the same conclusion is reached, namely the Stephani-D\c{a}browski model does not fit data, as long as one assumes the standard age of the Universe, 13.7 billion years. This does not rule out fitting the observational data if one allows the Universe to be older \cite{9905083}. This will, however, likely affect the thermal history of the Universe and the analysis regarding entropy in this work will need to be revised carefully.

\section{Conclusion}\label{VI}

Although the concordance $\Lambda$CDM model assumes a flat FLRW cosmology, which is homogeneous and isotropic, such an assumption might be an oversimplification. The effect of local inhomogeneity, if not carefully taken into account, may result in the wrong inferred value of the cosmological constant \cite{1104.0730}, or the impression that dark energy is evolving while in fact it is just a constant $\Lambda$ \cite{1006.4735}. Of course, it would have been a huge achievement for inhomogeneity cosmology if it could completely do away with $\Lambda$. This does not seem possible with the LTB models, which introduce inhomogeneity via $\rho=\rho(r,t)$. In this work, we investigated a specific model of Stephani cosmology -- the Stephani-D\c{a}browski model, which introduces instead, $p=p(r,t)$, and found that it too have difficulty satisfying observational constraints. This is despite the fact that, like some LTB models, the entropic bound required from the holographic principle remains satisfied. This implies that the holographic bound is not tight enough to serve as a good indicator for a viable cosmology. In other words, satisfying the holographic principle is a necessary but not sufficient condition. We also found that in Stephani universes, it is possible that the particle horizon is located far away from the apparent horizon, in contrast to a flat FLRW universe, in which these two horizons are close (i.e., their distances from the origin $r=0$ are of the same order of magnitude). It is interesting that in the Stephani-D\c{a}browski model, as the universe expands, the matter entropy decreases monotonically while the apparent horizon entropy increases, maintaining the overall increase in the total entropy as required by the generalized second law. This is consistent with our observation that the null energy condition (in fact, also the weak and dominant energy conditions) holds in the ``near-sheet'' of the spacetime geometry. 

To conclude, although Stephani cosmology is quite different from LTB cosmology, one still cannot do away with $\Lambda$, at least in the simplest models such as Stephani-D\c{a}browski's. We note that we have not yet utilized fully the features allowed by the Stephani solutions. In particular, off-centered observers can be considered \cite{1312.1567}.
Furthermore, in the model studied in this work, the spatial curvature is always negative, whereas spatial curvature is allowed to evolve and change sign in a generic Stephani cosmology. Whether such a cosmology could better accommodate observational data remains to be further examined. It would also be interesting to study how the matter entropy behaves in a more complicated Stephani model, and whether the holographic principle and the generalized second law could help to impose some theoretical constraints.

\begin{acknowledgments}
Yen Chin Ong would like to thank Brett McInnes, Projjwal Banerjee, and Fabien Nugier for useful discussions. He also thanks National Natural Science Foundation of China (No.11705162), Natural Science Foundation of Jiangsu Province (No.BK20170479), and China Postdoctoral Science Foundation (No.17Z102060070) for funding support. S. Sedigheh Hashemi would like to thank Shanghai Jiao Tong university for their warm hospitality during her visit. She also thanks Prof. Sh. Jalalzadeh for introducing the Stephani universes to her during her M.Sc. study. Bin Wang would like to acknowledge the support by National Basic Research Program of China (973 Program 2013CB834900) and National Natural Science Foundation of China. 
\end{acknowledgments}

\end{document}